\documentclass[prl,twocolumn,floatfix,showpacs ]{revtex4-1}
\UseRawInputEncoding
\usepackage{amsmath}
\usepackage{amssymb}
\usepackage{graphicx}% Include figure files
\usepackage{dcolumn}% Align table columns on the decimal point
\usepackage{bm}% bold math
\usepackage[colorlinks=true,linkcolor=blue,anchorcolor=blue, citecolor=cyan,urlcolor=cyan]{hyperref}% add hypertext capabilities
\usepackage[mathlines]{lineno}% Enable numbering of text and display math
\usepackage{ulem}
\usepackage{epstopdf}
\begin{document}
%========================================================
\title{Piezo-Hall effect in $\mathrm{RbCr_2Se_2O}$}
\author{San-Dong Guo$^{\dagger}$}
\email{sandongyuwang@163.com}
\affiliation{School of Electronic Engineering, Xi'an University of Posts and Telecommunications, Xi'an 710121, China}
\author{Shi-Hao Zhang}
\email{These authors contributed equally to this work.}
\affiliation{School of Physics and Electronics, Hunan University, Changsha 410082, China}
\author{Feng-Ren Fan}
\email{frfan@suda.edu.cn}
\affiliation{Institute for Quantum Science, School of Physical Science and Technology,  Soochow University, Suzhou 215006, China}
%=======================Abstract===============================================================
\begin{abstract}
The anomalous Hall effect (AHE) is of fundamental importance for understanding spin-orbit coupling and magnetic transport in magnetic materials. Meanwhile, strain can significantly modify material properties, and the piezoelectric effect stands as a well-known representative. The piezo-Hall effect combines the two physical ingredients of AHE and strain. Specifically, the piezo-Hall effect describes a physical process in which the magnetic space group (MSG) of a magnetic material prohibits a net magnetic moment and suppresses the AHE in the pristine state but transforms into a new MSG upon strain application that permits a net magnetic moment, thereby enabling the emergence of the AHE. The underlying mechanisms cover strain-induced magnetic electronic state transition and strain tuning of magnetization direction. We take the $d$-wave altermagnetic metal as an example to analyze the piezo-Hall effect in detail, with first-principles calculations carried out on the specific material
$\mathrm{RbCr_2Se_2O}$ to verify this scenario. A notable result is that under identical uniaxial strain applied along the
$x$ and $y$ directions, the anomalous Hall conductivity exhibits opposite signs but equal magnitudes. Similarly, the piezo-magneto-optical effect can also be proposed, and if strain tuning is replaced by electric-field modulation, the electro-Hall effect and electro-magneto-optical effect can be also put forward accordingly. Even temperature can serve as an external field to introduce the temperature Hall effect. Our work advances the understanding of the coupling between strain and physical properties.

\end{abstract}
\maketitle
%\tableofcontents
%========================Introduction===========================================================
\textcolor[rgb]{0.00,0.00,1.00}{\textbf{Introduction.---}}
Magnetism constitutes one of the most important research fields in condensed-matter physics and holds promising applications in information storage, spintronics, and quantum materials\cite{q1}. In particular, the discovery of unconventional magnetism has further stimulated intense research interest\cite{q2,q3}. Notably, altermagnetism, a newly identified magnetic state with alternating spin order, breaks the conventional dichotomy between ferromagnetism and antiferromagnetism\cite{q4}. In contrast to $PT$-antiferromagnets with the joint symmetry ($PT$) of space inversion symmetry ($P$) and time-reversal symmetry ($T$), altermagnet exhibits non-zero spin splitting in the absence of net magnetization, enabling exotic spin-dependent transport responses and offering new opportunities for developing dissipationless spintronic devices\cite{q5,q6,q7,q8,q9,q10,q11,q12,q13,q13-1,q14,q15}. Recently, another family of fully compensated ferrimagnet has also been revisited\cite{q16}; like altermagnets, these materials host nonvanishing spin polarization without macroscopic magnetic moments and thus constitute an alternative platform for exploring novel spin-transport phenomena\cite{q17,q18,q19,q20,q21}. Analogous to ferromagnets, the anomalous Hall effect (AHE) can emerge in such zero-net-moment systems when spin-orbit magnetism (SOM) is present\cite{q2}. As a fundamental spin-transport signature, the AHE serves as a powerful probe to characterize the magnetic and electronic structures of these unconventional magnets\cite{q2,q3,q22,q23,q24,q25}.

Strain can strongly couple to diverse physical properties of solids, giving rise to well-known symmetry-mediated phenomena such as the piezoelectric effect and the piezomagnetic effect\cite{q26,q27,q28,q29,q30}. The piezoelectric effect describes strain-induced electric polarization, while the piezomagnetic effect refers to the emergence of magnetic moments upon mechanical deformation. These examples demonstrate that mechanical strain acts as an effective symmetry knob for tailoring electronic and magnetic responses in functional materials. Very recently, the piezochiral effect has also been proposed\cite{q31}. Specifically, in a class of intrinsically achiral crystals, uniaxial strain lifts the compensation between locally counter-chiral fragments within each unit cell. The resulting handedness can be selected by altering the strain direction or switching between tensile and compressive deformation.
\begin{figure*}[t]
    \centering
    \includegraphics[width=0.90\textwidth]{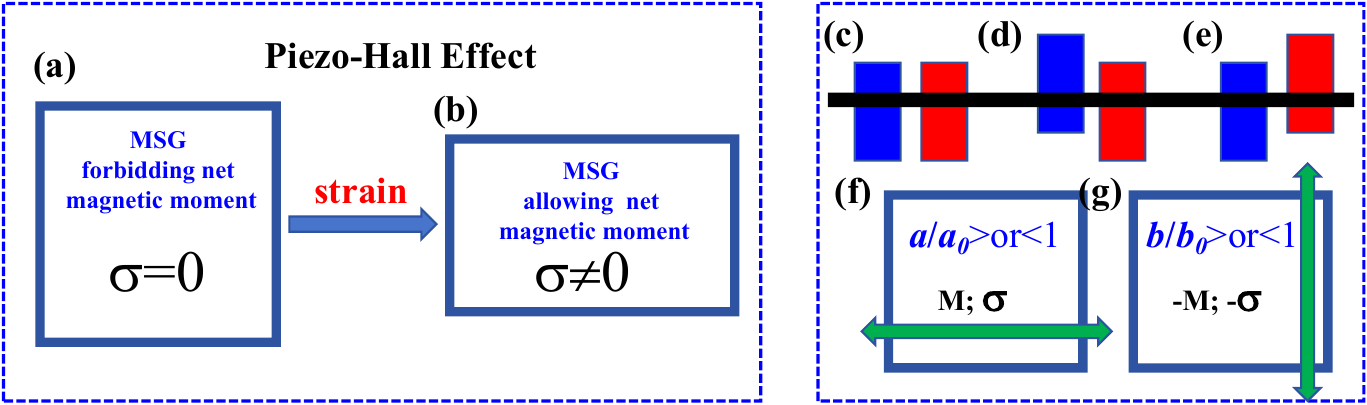}
    \caption{(Color online)  (a): for this magnetic material, the MSG forbids a net magnetic moment, suppresses the AHE, and yields zero AHC ($\sigma=0$). (b): upon strain application, the MSG of this material becomes compatible with a net magnetic moment, enabling the AHE and giving rise to a non‑zero AHC   ($\sigma\neq0$). The transition from (a) to (b) is referred to as the piezo-Hall effect.   (c, d, e): the schematic diagram of band shifts under an applied uniaxial strain, including unstrained, $x$-direction strained, and $y$-direction strained cases. (f, g): the sign variations of total magnetic moment ($M$) and AHC with strain ($a/a_0$ or $b/b_0$) applied from the $x$ to $y$ direction. In (c, d ,e), the blue and red  represent spin-up and spin-down bands, respectively, while the horizontal black line denotes the Fermi level. In (f, g), the green arrows denote the strain direction.}\label{a}
   \end{figure*}

\begin{figure}[t]
    \centering
    \includegraphics[width=0.38\textwidth]{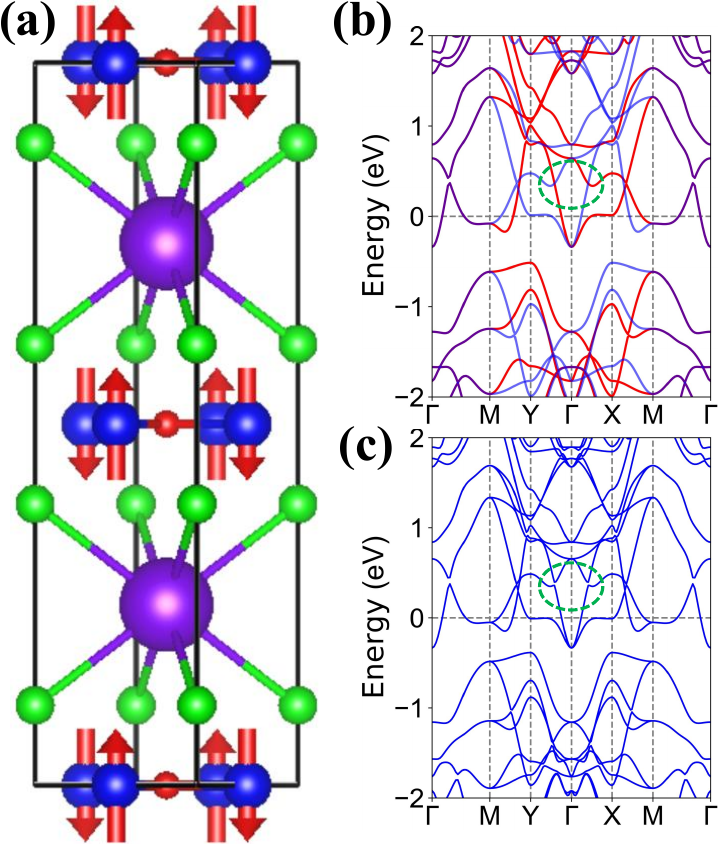}
    \caption{(Color online) For $\mathrm{RbCr_2Se_2O}$, (a): the crystal structure with 1$\times$1$\times$2 supercell. (b): the strain-free spin-polarized energy  band structures without considering SOC. (c): the strain-free  energy band structures with SOC. In (a), the purple, blue, green, and red spheres represent Rb, Cr, Se, and O atoms, respectively. The red arrows represent spin directions, illustrating a C-type AFM configuration. Flipping the $\mathrm{N\acute{e}el}$ vector of one Cr atomic layer yields a G-type AFM state. In (b), the blue, red, and purple denote the spin-up, spin-down, and spin-degenerate bands. In (b, c), the green dashed circle highlights the small band gap opened upon the inclusion of SOC.}\label{b}
   \end{figure}

  Can strain couple to the AHE in zero-net-moment magnetic materials?  Recently, a distinct piezo-Hall effect different from those previously reported has been identified\cite{ljw}. Simply put, the AHE vanishes in the absence of strain and emerges upon strain application. Even so, more detailed investigations and calculations remain to be implemented. Herein, we demonstrate the feasibility of the piezo-Hall effect by using concrete material example $\mathrm{RbCr_2Se_2O}$ and show that the sign of the anomalous Hall conductivity (AHC) can be reversed by tuning the strain direction. By analogy, the piezo-magneto-optical effect can also be envisioned. If strain tuning is substituted by electric-field modulation, analogous concepts including the electro-Hall effect and electro-magneto-optical effect may likewise be proposed. More generally, whenever strain enables a physical quantity or response to emerge from complete suppression, one may define a corresponding piezo- effect. Furthermore, temperature can also act as the external field. As an example, the high-temperature structural phase possesses a symmetry that prohibits the AHE and related effects, whereas the low-temperature phase allows these effects, which likewise gives rise to the temperature Hall effect, and so on.

\begin{figure*}[t]
    \includegraphics[width=0.9\textwidth]{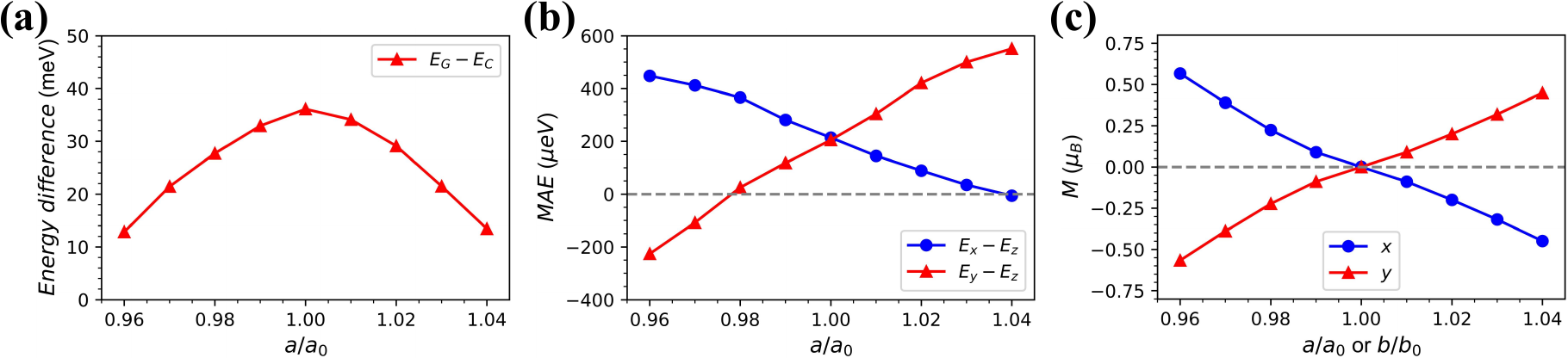}
    \caption{(Color online)  For $\mathrm{RbCr_2Se_2O}$ with 1$\times$1$\times$2 supercell, (a): the energy difference between G-type and C-type magnetic configurations as a function of $a/a_0$, and the C-type is taken as the reference and assigned a zero value. (b): the MAE as a function of $a/a_0$, where the out-of-plane magnetization direction is set as the reference with a zero value. (c): the total magnetic moment as a function of $a/a_0$ or  $b/b_0$, including two cases of uniaxial strain along the $x$ and $y$ directions.}
    \label{c}
\end{figure*}

\textcolor[rgb]{0.00,0.00,1.00}{\textbf{Piezo-Hall effect.---}}
Whether the AHE can arise in a magnetic material depends on whether its magnetic space group (MSG) permits the generation of a net magnetic moment\cite{q2,fs}.
First, the MSG of a magnetic material must forbid the generation of a net magnetic moment, thereby suppressing the AHE and producing the zero AHC (\autoref{a} (a)).  For example, $PT$-antiferromagnets and some altermagnets without SOM can satisfy the above conditions\cite{q2,q3}. It should be noted that altermagnets are defined in the non-relativistic limit. For the same material, when spin-orbit coupling (SOC) is taken into account, different magnetization directions will give rise to distinct MSG\cite{q2,q3}, which directly determines whether the AHE can emerge.
Next, upon applying strain, the MSG of this material transforms into one that permits a net magnetic moment. Consequently, the AHE becomes allowed, giving rise to a nonzero AHC (\autoref{a} (b)). For instance,  the resulting ferrimagnets, fully compensated ferrimagnets and some altermagnets with SOM can satisfy the mentioned conditions\cite{q2}.
The strain-induced AHE can be referred to as the piezo-Hall effect (\autoref{a} (a) to  \autoref{a} (b)). We present several examples: strain can induce a transition from $PT$-antiferromagnetism or SOM-free altermagnetism to fully compensated ferrimagnetism or ferrimagnetism via symmetry breaking; alternatively, strain can modulate the magnetization direction of altermagnet to switch the system from SOM-free to SOM-active. All these mechanisms can realize the piezo-Hall effect.

In the following, we take the $d$-wave altermagnetic metal as an example to analyze the  piezo-Hall effect  in detail.
In the absence of strain, the system exhibits no AHE, obeys symmetry $[C_2||C_4]$ ($C_2$ denotes spin-flip in spin space, while $C_4$ represents the four-fold rotation operation in real space.), and has a zero total magnetic moment. The two schematic  alternating energy bands along with the strain-applied cases are shown in \autoref{a} (c, d, e). For the unstrained case, spin-up and spin-down electron populations are balanced (\autoref{a} (c)), corresponding to zero net magnetic moment. When uniaxial strain is applied, the system generally becomes a ferrimagnetic metal due to broken $[C_2||C_4]$ symmetry. Upon applying  strain along the $x$ direction, the spin-up bands are supposed to shift upward relative to the spin-down bands  (\autoref{a} (d)). Fewer spin-up electrons than spin-down electrons lead to an  existing magnetic moment, and the AHC emerges naturally (\autoref{a} (f)).
When strain is along the $y$ direction, the spin-down bands shift upward with respect to the spin-up bands. Fewer spin-down electrons yield a opposite magnetic moment, accompanied by a sign change of the AHC (\autoref{a} (g)).

We will take the recently predicted robust $d$-wave altermagnet $\mathrm{RbCr_2Se_2O}$\cite{q14} as an example to verify our proposal, and the polycrystals of this magnetic material have been experimentally synthesized\cite{q32}.

\textcolor[rgb]{0.00,0.00,1.00}{\textbf{Computational detail.---}}
The spin-polarized  calculations are  carried out  within density functional theory (DFT)\cite{1}
using the Vienna ab initio simulation package (VASP)\cite{pv1,pv2,pv3} by using the projector augmented-wave (PAW) method. The generalized gradient approximation (GGA) proposed by
Perdew, Burke, and Ernzerhof (PBE)\cite{pbe} is adopted  as  the exchange-correlation functional. The  kinetic energy cutoff of 500 eV,  total energy  convergence criterion of  $10^{-8}$ eV and  force convergence criterion of 0.001 $\mathrm{eV\cdot{\AA}^{-1}}$ are used  to obtain reliable results.
A 14$\times$14$\times$8 Monkhorst-Pack $k$-point meshes are adopted to sample the Brillouin zone (BZ)  for structural relaxation and electronic structure calculations.
The maximally localized Wannier functions  are constructed using the Wannier90 code\cite{w1}. The intrinsic AHC is  calculated by integrating the Berry curvature over the BZ with a dense 256$\times$256$\times$128 $k$-point meshes  via the built-in Berry postprocessing module.

\textcolor[rgb]{0.00,0.00,1.00}{\textbf{Material realization.---}}
 The  quasi-two-dimensional $\mathrm{RbCr_2Se_2O}$ has been predicted to be a $d$-wave altermagnet with a C-type antiferromagnetic (AFM) configuration\cite{q14}. Its lattice structure is identical to those of the extensively studied $\mathrm{KV_2Se_2O}$,  $\mathrm{Rb_{1-\delta}V_2Te_2O}$  and $\mathrm{Cs_{1-\delta}V_2Te_2O}$\cite{q12,q13,q13-1}, and can be regarded as Rb atoms intercalated between two  $\mathrm{Cr_2Se_2O}$  monolayers (\autoref{b} (a)).  Recent experiments have confirmed that $\mathrm{KV_2Se_2O}$,  $\mathrm{Rb_{1-\delta}V_2Te_2O}$  and $\mathrm{Cs_{1-\delta}V_2Te_2O}$ all adopt the G-type AFM configuration\cite{q13-1,kkk,kkkk}, which corresponds to the so-called hidden altermagnetism with global $PT$ symmetry\cite{q15}.  Theoretically, the energy difference between their C-type and G-type configurations is small. By contrast, for $\mathrm{RbCr_2Se_2O}$, the energy of the G-type configuration is distinctly higher than that of the C-type one\cite{q14}, ensuring robust altermagnetism.
The MSG of a material depends on its magnetization direction, which in turn governs its associated physical properties. When the magnetization points out-of-plane, $\mathrm{RbCr_2Se_2O}$  adopts  MSG of $P4^\prime/mm^\prime m$, which forbids the emergence of the AHC.
Nevertheless, when the magnetization direction lies in-plane along the $x$ or $y$ direction, the MSG becomes
$Pm^\prime m^\prime m$, which permits the occurrence of the AHE. $\mathrm{RbCr_2Se_2O}$ has been confirmed to exhibit out-of-plane magnetization\cite{q14,q32}, which guarantees the feasibility of verifying our proposal. Assuming
$\mathrm{RbCr_2Se_2O}$ takes on a G-type magnetic configuration,
$PT$  symmetry forbids the AHE for both out-of-plane and in-plane magnetizations. Hence, for a given material, both magnetic order and magnetization direction affect the emergence of the AHE.
\begin{figure*}[t]
    \centering
    \includegraphics[width=0.96\textwidth]{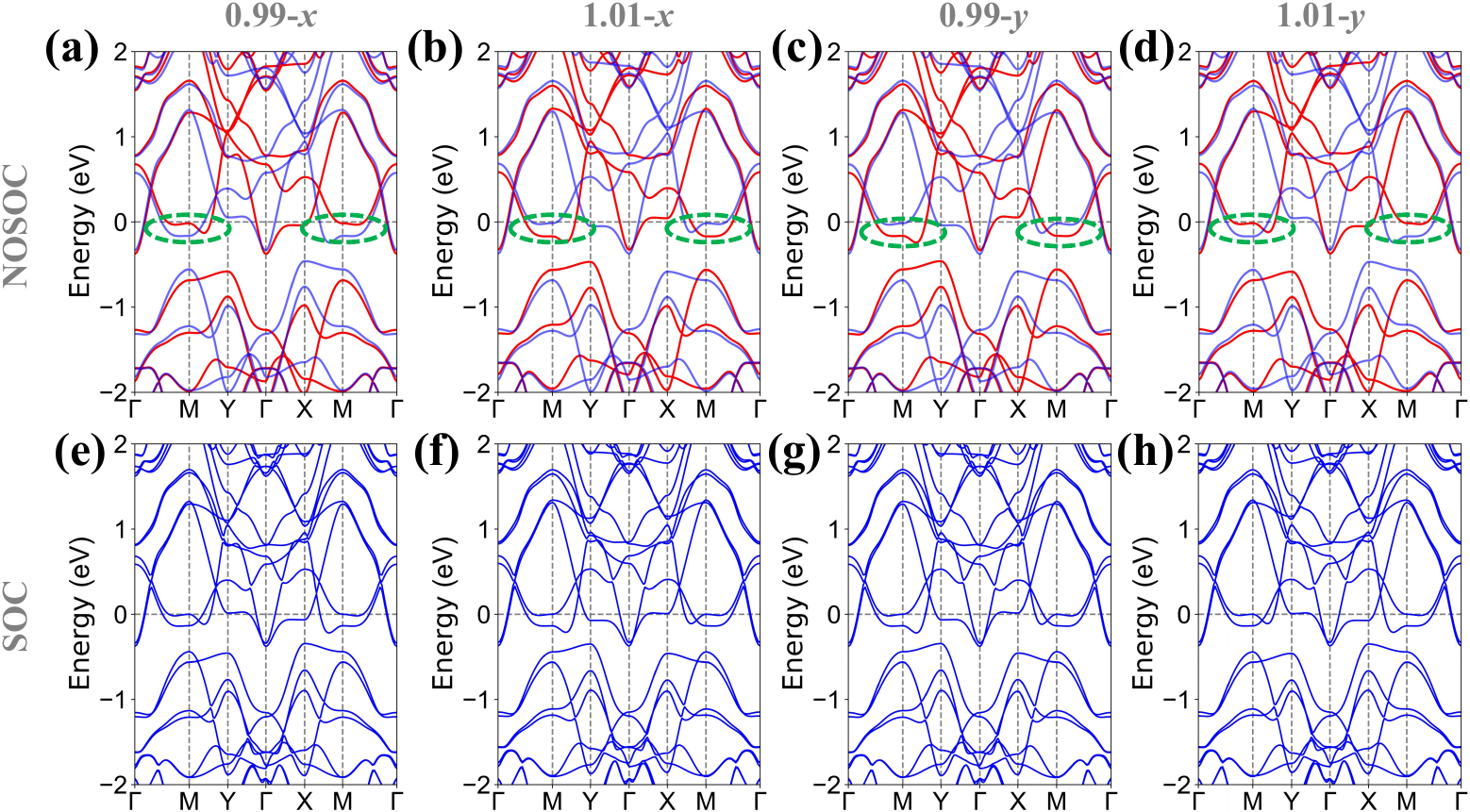}
    \caption{(Color online) For $\mathrm{RbCr_2Se_2O}$,  (a, b, c, d): the  spin-polarized energy  band structures without considering SOC at uniaxial strain 0.99 and 1.01  along the $x$ and $y$ directions. (e, f, g, h): the  energy band structures with SOC at uniaxial strain 0.99 and 1.01  along the $x$ and $y$ directions. In (a, b, c, d),  the blue and red denote the spin-up and  spin-down bands, and the green dashed circles highlight the representative  alternating energy
bands.}\label{d}
\end{figure*}

The C-type magnetic structure of $\mathrm{RbCr_2Se_2O}$ possesses the $[C_2||C_4]$ symmetry, displaying $d$-wave altermagnetism in its band structure, as shown in \autoref{b} (b). It is clearly seen that spin degeneracy exists along the
$\Gamma$-M path, whereas alternating spin splitting occurs along  the M-Y-$\Gamma$ and M-X-$\Gamma$
 paths. It can also be clearly observed that all bands at the $\Gamma$ point exhibit spin degeneracy, satisfying the fundamental requirement for altermagnetism\cite{q4}.
More importantly, $\mathrm{RbCr_2Se_2O}$ is metallic, which provides a prerequisite for directly generating the AHE without the need for carrier doping.
Upon the inclusion of SOC, the overall band profile remains unchanged, while small band gaps open at several positions (see \autoref{b} (c)).

In the following, we employ uniaxial strain to break the symmetry of $\mathrm{RbCr_2Se_2O}$, thereby altering the MSG and allowing the AHE to emerge. Strain is characterized by $a/a_0$ or $b/b_0$ along the $x$ or $y$ direction.
Upon applying uniaxial strain, the MSG of
$\mathrm{RbCr_2Se_2O}$ becomes $Pm^\prime m^\prime m$ even for the C-type magnetic structure with out-of-plane magnetization, enabling the occurrence of the AHE. Naturally, the AHE is allowed under in-plane magnetization, with or without strain.
For the G-type magnetic configuration, the AHE remains forbidden even under uniaxial strain, since
$PT$ symmetry is preserved.

Therefore, we first calculate the energy difference between the G-type and C-type magnetic configurations to confirm that the C-type phase is the ground state, so that the AHE can be realized upon applying uniaxial strain. \autoref{c} (a) shows that the C-type configuration remains the ground state within the considered strain range. Naturally, since the energy does not depend on strain direction, the results are identical when strain is applied along either the
$x$  or $y$ direction. We then consider the evolution of the magnetization direction with strain.   The magnetization direction can be determined by the magnetic anisotropy energy (MAE), which is defined as  energy difference $E_x$-$E_z$ and $E_y$-$E_z$, where $E_x$/$E_y/$$E_z$ is the energy per magnetic cell  when the magnetization is along the $x$/$y$/$z$ direction.
The corresponding MAE is presented in \autoref{c} (b). It should be pointed out that the only difference between the
$x$- and $y$-direction ($a/a_0$ and $b/b_0$) cases is that the two curves are swapped. The out-of-plane magnetization can be maintained when the strain lies between 0.98 and 1.04. Naturally, whether the magnetization direction changes under uniaxial strain has no impact on our strain-induced AHE, since the AHE emerges irrespective of the magnetization direction once strain is applied. Nevertheless, it would be more desirable for us to study the strain-induced AHE with the magnetization direction kept fixed.

Upon applying uniaxial strain, the $C_4$ symmetry is broken, and there exist no symmetry-related connections between the spin-opposite magnetic atoms. For a metallic system, it may turn into a fully compensated ferrimagnetic metal or a ferrimagnetic metal\cite{q20,gsd}.
In general, a semimetal will transform into a fully compensated ferrimagnetic metal\cite{q20}, with the total magnetic moment remaining zero.
For $\mathrm{RbCr_2Se_2O}$, it should evolve into a ferrimagnetic metal with applied strain.
The dependence of the total magnetic moment on uniaxial strain  ($a/a_0$ and $b/b_0$) along the $x$ and $y$ directions is presented in \autoref{c} (c).
For uniaxial strain along the same direction, compressive and tensile strains yield opposite sign of the magnetic moment. For the same magnitude of strain applied along different direction, the magnetic moments are also opposite in sign but identical in magnitude. The sign change of the magnetic moment is consistent with our previous discussion (see \autoref{a}).

\begin{figure}[t]
    \centering
    \includegraphics[width=0.48\textwidth]{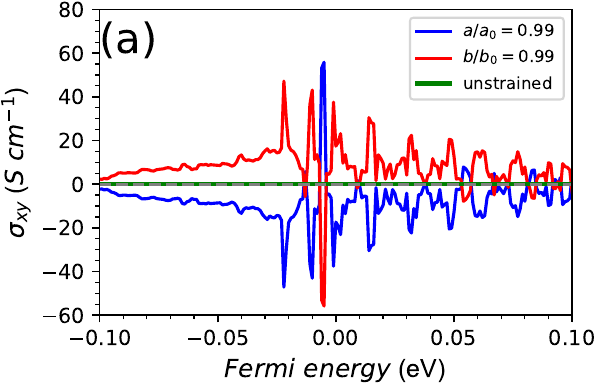}
      \includegraphics[width=0.48\textwidth]{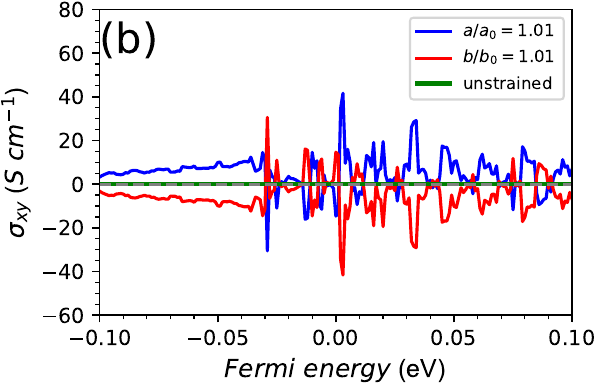}
    \caption{(Color online)  For $\mathrm{RbCr_2Se_2O}$ at 0.99 (a) and 1.01 (b) strains along $x$ ($a/a_0$) and $y$ ($b/b_0$) directions together with unstrained case,  the  AHC as a function of Fermi energy.}\label{e}
   \end{figure}

The  spin-polarized energy  band structures at uniaxial strain 0.99 and 1.01  along the $x$ and $y$ directions  without  SOC  and with SOC are plotted in \autoref{d}.
For strain applied along the same direction, the band structure under strain 0.99 approximately swaps paths M-Y-$\Gamma$ and M-X-$\Gamma$ of the 1.01 case and flips the band spins. For the same magnitude of strain, the energy bands along the $x$-direction can be exactly obtained by swapping paths M-Y-$\Gamma$ and M-X-$\Gamma$ and flipping the band spins of the energy bands along the $y$-direction.
These relations can also be used to understand how the total magnetic moment correlates with the strain type (tensile or compressive) as well as the strain direction.
We select the representative  alternating energy
bands along paths Y-M-$\Gamma$ and X-M-$\Gamma$. As can be seen in \autoref{d}, the variations of   alternating energy
bands under different strains are consistent with our previous analysis (see \autoref{a} (d, e)). Upon including SOC, the overall trend and profile resemble the SOC-free case, with energy gaps opening only in some band regions.

Finally, we calculate the AHC of  $\mathrm{RbCr_2Se_2O}$ under strains of 0.99 and 1.01 along the $x$  and
$y$  directions together with the zero-strain case, which are plotted in \autoref{e}. The calculation results show that the AHC of $\mathrm{RbCr_2Se_2O}$ with out-of-plane magnetization is indeed zero under zero strain, which is consistent with symmetry analysis. It is found that uniaxial strain can indeed induce the AHC, namely the piezo-Hall effect.
Under the same compressive or tensile strain, the signs of the AHC along the
$x$ and $y$ directions are opposite, but their magnitudes are equal. This is consistent with the observation that for the same magnitude of strain applied along different direction, the magnetic moments are also opposite in sign but identical in magnitude. Another noteworthy point is that for the same direction, the AHC exhibits a similar overall profile under strains of 0.99 and 1.01, yet shows an opposite trend in sign. This arises from the opposite signs of the magnetic moments in these two cases. In a word, a small strain can indeed induce the AHE in $\mathrm{RbCr_2Se_2O}$, which verifies our proposal.

\begin{figure}[t]
    \centering
    \includegraphics[width=0.48\textwidth]{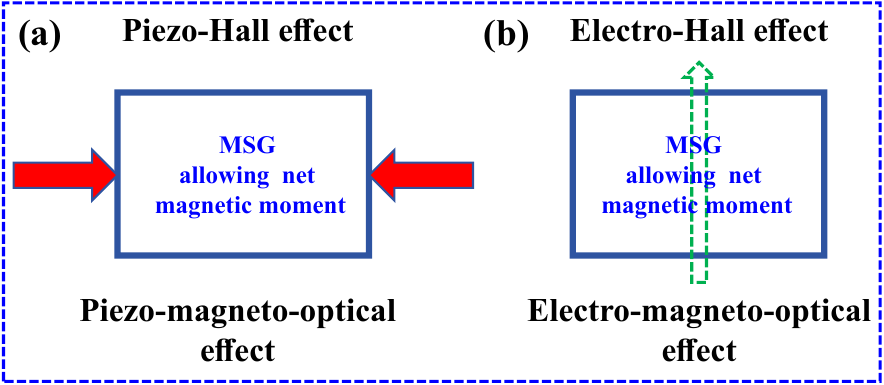}
    \caption{(Color online)  (a): under applied strain, the MSG allows for net-magnetic-moment formation, enabling the emergence of both the piezo-Hall effect and piezo-magneto-optical effect. (b): under an applied electric field, the MSG allows for net-magnetic-moment formation, enabling the emergence of both the electro-Hall effect and electro-magneto-optical effect. In (a), red arrows denote strain; in (b), green arrows represent the electric field.}\label{f}
   \end{figure}

\textcolor[rgb]{0.00,0.00,1.00}{\textbf{Discussion and Conclusion.---}}
Besides inducing the AHE by breaking the system symmetry via strain, the AHE can also be triggered by an electric field, which is referred to as the electro-Hall effect. Monolayer $\mathrm{Cr_2O}$ has been predicted to be an out-of-plane $d$-wave altermagnetic metal\cite{m1}.
It also serves as a candidate material for realizing the piezo-Hall effect. In the absence of strain, the MSG of $\mathrm{Cr_2O}$ for out-of-plane magnetization is $P-4^\prime m^\prime 2$, which forbids the AHE. Upon applying uniaxial strain, the MSG transforms to $Pm^\prime m^\prime 2$, allowing the AHE to occur.
The $\mathrm{Cr_2O}$ adopts an A-type AFM configuration. When an out-of-plane electric field is applied, the upper and lower Cr atomic layers become nonequivalent\cite{q20}, and the MSG of $\mathrm{Cr_2O}$ also  transforms to $Pm^\prime m^\prime 2$, which allows the occurrence of the AHE.
Hence, the $\mathrm{Cr_2O}$ also serves as a candidate material for  observing the electro-Hall effect.
The electro-Hall effect  can also be realized in A-type $PT$-antiferromagnets, such as experimentally synthesized
monolayer MnSe\cite{m2}. An A-type $PT$-antiferromagnet can  also be constructed by stacking ferromagnetic monolayers into bilayers\cite{m30},  such as experimentally synthesized bilayer  $\mathrm{CrPS_4}$\cite{m3}. Since $PT$-antiferromagnetism forbids the occurrence of the AHE, applying an out-of-plane electric field renders the two magnetic sublayers nonequivalent, driving the system into a ferrimagnetic or fully compensated ferrimagnetic state that permits the AHE.
The AHE and magneto-optical effects (Faraday and Kerr effects) share similar physical origin: they are related to the Berry curvature of electrons in reciprocal space, and they are subject to identical symmetry constraints\cite{q2,fs}.  Therefore, one can also introduce the piezo-magneto-optical effect and electro-magneto-optical effect. The related strain- and electric-field-induced AHE and magneto-optical effects are summarized in \autoref{f}.

 More generally, a piezo- effect can be defined whenever strain drives a physical quantity or phenomenon from absence to emergence.
For instance, strain can couple with the anomalous Nernst effect and photovoltaic effect, yielding the so-called piezo-Nernst effect, piezophotovoltaic effect, and so forth. Furthermore, the external field here can also be temperature. For instance, the symmetry of the high-temperature structural phase forbids effects such as the AHE, while the symmetry of the low-temperature structural phase permits them. Similarly, a temperature Hall effect, and so on, can be introduced.

In summary, the piezo-Hall effect combines  strain modulation and anomalous Hall transport. The underlying symmetry mechanism relies on strain-induced transformation of the MSG: the pristine MSG prohibits a net magnetic moment and suppresses the AHE, whereas strain alters the MSG to enable net magnetization and the AHE. Using the
$d$-wave altermagnetic metal as an example, we explore this effect and verify it through first-principles calculations of
$\mathrm{RbCr_2Se_2O}$. Our results show that the AHC exhibits opposite signs but equal magnitudes under identical uniaxial strain along the
$x$ and $y$ directions.  These findings can also be verified in other members of the $\mathrm{XCr_2Y_2O}$  (X=K, Rb, Cs; Y=S, Se, Te) family\cite{q14}; in particular, compound $\mathrm{CsCr_2S_2O}$ has also recently been experimentally synthesized as a high-temperature phase\cite{m4}.
Subsequently, examples of strain-tunable magnetization orientation can be sought to verify the piezo-Hall effect. Our work may stimulate relevant experiments and further enrich the family of piezo-related effects.

\begin{acknowledgments}
This work is supported by  the National Natural Science Foundation of China  (Grant No.12674079 and Grant No.12504217). We are grateful to Shanxi Supercomputing Center of China, and the calculations were performed on TianHe-2.
\end{acknowledgments}

\end{document}